\documentclass{article}
\usepackage[dvips]{graphics}
\usepackage[intlimits]{amsmath}
\usepackage{amssymb}
\DeclareMathOperator{\dlim}{\mathrm{d-lim}} 
\DeclareMathOperator{\tr}{\mathrm{Tr}} 
\begin{document}
\title{Mean-field heat capacity of dilute magnetic alloys}
\author{J. Ma\'ckowiak\footnote{Email address: \texttt{ferm92@fizyka.umk.pl}}\\ 
\textit{Universytet M. Kopernika, Instytut Fizyki, Grudzi\c adzka 5,}\\
\textit{87-100 Toru\'n, Poland}}
\maketitle

PACS: 72.15.Qm, 
	75.10.Jm, 
	75.30.Hx 

\begin{abstract}
Using an asymptotic solution of the~$M$-impurity thermodynamics of a dilute \mbox{s-d}~system, the impurity
energy and impurity heat capacity~$\Delta C(T)$ are derived for dilute magnetic alloys with spin~$1/2$ and
spin~$3/2$ impurities. The parameters which enter~$\Delta C$ are adjusted to fit experimental data on
impurity heat capacity of~$\mathrm{CuCr}$ and $\left( \mathrm{La}_{1-x}\mathrm{Ce}_x\right)\mathrm{Al}_2$.
Agreement is satisfactory for~$\mathrm{CuCr}$, at temperatures below~$1\mathrm{K}$, and good for~$\left(
\mathrm{La}_{1-x}\mathrm{Ce}_x\right)\mathrm{Al}_2$. The magnitude of theoretical~$\Delta C(T)$ agrees with
experiment and does not require scaling as in previous \mbox{s-d} theories. Nonlinear dependence of $\Delta
C(T)$ on impurity concentration has been accounted for the first time.
\end{abstract}

\section{Introduction}
\label{sec1}
Existing theories of anomalous thermal behaviour of dilute magnetic alloys (DMA) 
(e.\ g.\ Refs.~\cite{kondo,bloom,wilson,rajan,andrei,filyov,wiegmann,hewson,m1}) have provided only partial 
quantative explanation of these anomalies. Most of these investigations were restricted to an 
\mbox{s-d}~system consisting of a single magnetic impurity interacting with the electron gas. 

Significant progress was made by \mbox{Andrei}~et~al.~\cite{rajan,andrei}, who solved the 
\mbox{s-d}~thermodynamics for an \mbox{s-d}~system with impurities treated as indistinguishable particles. 
Their heat capacity and magnetization curves, after scaling by an adjusted factor, provide a good fit to 
experimental data on~$\left(\mathrm{La}_{1-x}\mathrm{Ce}_x\right)\mathrm{Al}_2$. 

The disadvantage of almost all existing theories of DMA is linear dependence of resulting thermodynamic
functions on impurity concentration $c$. The dependence of DMA experimental data on $c$ is more complex, e.g.
the temperature at which DMA resistivity minimum occurs is proportional to $c^{1/5}$~\cite{kondo} and 
normalized DMA impurity heat capacity $\Delta C /c$ is not constant in $c$~\cite{triplett}.

A different approach to the thermodynamics of DMA, in which the termodynamic functions do not require
rescaling and depend on $c$ nonlinearly, was proposed by the author in Ref.~\cite{m2}. The starting point in 
this approach is the first quantization \mbox{s-d}~Hamiltonian~$H_{\mathrm{I}}^{(n,M)}$ describing 
$n$~electrons and $M$~distinguishable, arbitrarily positioned impurities with single-component spins 
interacting via a locally smeared \mbox{s-d} interaction~$U(\vec{R}_{\alpha}-\vec{r}_i) S_{z\alpha} 
\sigma_{zi}$:
\begin{equation}
	H_{\mathrm{I}}^{(n,M)}=A^{(n)}\left( H_0^{(n)} + g^2 \sum_{\alpha=1}^M \sum_{i=1}^n
	U(\vec{R}_{\alpha}-\vec{r}_i)S_{z\alpha}\sigma_{zi} \right)\text{,}	
	\label{H_I}
\end{equation}
where~$A^{(n)}$ denotes the antisymmetrizer with respect to electron variables with indices~$i=1,\ldots ,n$, 
\begin{equation}
	H_0^{(n)}=-\frac{\hslash^2}{2m}\sum_{i=1}^n \mathit{\Delta}_i \text{.}
	\label{H0}
\end{equation}
$\vec{R}_{\alpha}$ denotes the position of~$\alpha$th impurity, $S_{z\alpha}$~its spin operator and
$\vec{r}_i$,~$\sigma_{zi}$ the corresponding quantities of the $i$th electron. The considerations of 
Ref.~\cite{m2} remain valid for any sufficiently regular function~$U$ depending on $|\vec{R}_{\alpha}-
\vec{r}_i|$.

It was shown in Ref.~\cite{m2}, that in the dilute limit of small $c$ ($\dlim$) the \mbox{s-d} interaction is 
separable and the system's free~energy per electron $f(H_{\mathrm{I}}^{(n,M)},\beta)$ is equal asymptotically 
to that of $n$~noninteracting electrons with the 1-electron Hamiltonian
\begin{subequations}
\begin{equation}
	h_{\mathrm{e}}^{(1,M)}(\xi,\eta)=\widetilde{h}_{\mathrm{e}}^{(1,M)}(\xi,\eta)+\frac{1}{2}M(\xi^2-\eta^2)\mathbb{I}\text{,}
	\label{h_e}
\end{equation}
where
\begin{equation}
	\widetilde{h}_{\mathrm{e}}^{(1,M)}(\xi,\eta)=H_0^{(1)}-g\sqrt{n}(\xi-\eta)\sum_{\alpha=1}^M
	U_{\alpha}^{(1)}\sigma_z^{(1)}
	\label{h_e_tilde}
\end{equation}
\end{subequations}
and $M$~impurities described by the Hamiltonian
\begin{equation}
	h_{\mathrm{imp}}^{(M)}(\xi)=g\sqrt{n}\xi\sum_{\alpha=1}^M S_{z\alpha}+\frac{1}{2} g^2 \sum_{\alpha=1}^M
	S_{z\alpha}^2\text{.}
	\label{h_imp}
\end{equation}
$U_{\alpha}^{(1)}$~in Eq.~\eqref{h_e_tilde} is the multiplication operator by~$U(\vec{R}_{\alpha}-
\vec{r}_i)$ and $\eta(\xi)=\xi-f_2(\xi)$, with
\begin{equation}
	f_2(\xi)=-\frac{g}{\sqrt{n}}\left\langle S_z \right\rangle_{h_{\mathrm{imp}}^{(1)}}\quad \text{,} 
	\quad \left\langle B \right\rangle_{h} :=  \frac{\tr(B\mathrm{e}^{-\beta h})}{\tr\mathrm{e}^{-\beta h}}\text{,}
	\label{f2}
\end{equation}
whereas~$\xi$ is the solution of the equation
\begin{equation}
	\xi=f_1\left(f_2(\xi)\right)+f_2(\xi)
	\label{xi}
\end{equation}
with
\begin{equation}
	f_1(\xi)=g\sqrt{n}\left\langle \Gamma_1^n U_{\alpha}^{(1)}\sigma_z^{(1)}\right\rangle_{n\Gamma_1^n
	\widetilde{h}_{\mathrm{e}}^{(1,M)}(\xi,0)}\text{,}
	\label{f1}
\end{equation}
\begin{equation*}
	\Gamma_1^n B^{(1)}:= A^{(n)}\left( B^{(1)}\otimes \mathbb{I}^{(n-1)}\right)A^{(n)}\text{,}
	\label{gamma}
\end{equation*}
which minimizes the mean-field free~energy per electron~$f(h^{(n,M)},\beta)$, where
\begin{equation}
	h^{(n,M)}(\xi,\eta)=h_{\mathrm{e}}^{(n,M)}(\xi,\eta)+h_{\mathrm{imp}}^{(M)}(\xi)\text{.}
	\label{h_nm}
\end{equation}
Asymptotic equivalence of~$H_{\mathrm{I}}^{(n,M)}$ and~$h^{(n,M)}$ is expressed by the equality
\begin{equation}
	\lim_{n\to\infty} \dlim f(H_{\mathrm{I}}^{(n,M)},\beta)=\lim_{n\to\infty}\dlim f(h^{(n,M)},\beta) 
	=\lim_{n\to\infty}f(A^{(n)} H_0^{(n)},\beta)\text{.} 
	\label{asympt}
\end{equation}

The question arises, to what extent is the \mbox{mean-field} \mbox{s-d}~Hamiltonian~$h^{(n,M)}$ capable of 
providing a reliable theory of DMA thermal behaviour. Here a partial answer to this question is found by 
showing that impurity heat capacity curves of two~$\mathrm{CuCr}$ alloys~\cite{triplett} with $c=21.7
\,\mathrm{ppm}\text{,} \, 51\, \mathrm{ppm}$ and
$(\mathrm{La}_{1-x}\mathrm{Ce}_x)\mathrm{Al}_2$~\cite{bader} with $x=0.0064$ can be satisfactorily explained
in terms of $h^{(n,M)}$ thermodynamics.

To this end, the 1-particle equilibrium density operator~$\rho^{(1)}$ of a quantum gas, in a field of randomly
positioned wells, is first analysed in Section~\ref{2}. Such operator appears, e.g.\ in the equation for the
fugacity~$z$:
\begin{equation}
	\tr z\rho^{(1)}\left( \mathbb{I}+z\rho^{(1)}\right)^{-1}=n
	\label{z}
\end{equation}
where
\begin{equation*}
	\rho^{(1)}:= \exp \left[ -\beta \widetilde{h}_{\mathrm{e}}^{(1,M)}(\xi,0)\right]
\end{equation*}
It is shown that, in the low-temperature r\'egime, such a gas behaves effectively like a system of free particles at a temperature higher than
that of the real system. 

In Section~\ref{3} the \mbox{mean-field} impurity energy~$\Delta U_{\mathrm{s-d}}$ of the \mbox{s-d} system, relative to that of the free electron gas, is derived for impurity spins~$1/2$,~$3/2$. Here the crucial 
question is the form of~$f_1$. An approximate \mbox{Sommerfeld}-type expansion of~$f_1(\xi)$ is found, which 
is subsequently applied in the simplest truncated form, viz.,~$f_1(\xi)=b_0+b_1 \xi$, with~$b_0, b_1 \in 
\mathbb{R}^1$. As a consequence, $\Delta U_{\mathrm{s-d}}$~depends on five parameters:~$b_0$, $b_1$, $g$, $M$, 
$\Delta T$, $M$ denoting the number of impurities in the considered \mbox{s-d} subsystem of the molar 
\mbox{s-d} system and $\Delta T>0$~the shift in temperature scale due to random interactions unaccounted for 
by the \mbox{s-d}~Hamiltonian~\eqref{H_I}.

The expression for the \mbox{mean-field} heat capacity
\begin{equation}
	\Delta C(T)=\frac{\mathrm{d} \Delta U_{\mathrm{s-d}}}{\mathrm{d} T}
	\label{deltaC}
\end{equation}
was calculated analytically and the parameters~$b_0$, $b_1$, $g$, $M$, $\Delta T$ adjusted to obtain
directly the best possible fit of \mbox{mean-field}~$\Delta C(T)/c$ with experimental data, without
additional rescaling procedures.  

Variation of~$\Delta C/c$ with $c$~\cite{triplett} requires different values of $b_0$, $b_1$, $g$, $M$,
$\Delta T$ for different impurity concentrations. In this manner, for the first time, nonlinearity
of~$\Delta C(T)/c$ with respect to~$c$ has been accounted for. 
\section{1-particle density operator of a quantum gas in a field of randomly positioned wells}\label{2}
The Hamiltonian~$H_{\mathrm{I}}^{(n,M)}$ gives only a simplified account of the interactions present in an 
\mbox{s-d}~system. It does not contain terms representing the \mbox{Coulomb} interactions between electrons 
and the screened \mbox{Coulomb} pseudo-potential $\sum_{\alpha}U_{C\alpha}$ of each electron in the field of 
impurities. In order to investigate the effect of such potential on the electron gas in a DMA, let us examine 
the \mbox{1-particle} density operator of a gas of negatively charged particles interacting with randomly 
positioned positively charged ions in a region~$\Lambda$. 

Let
\begin{equation}
	h^{(1)}=-\frac{\hslash^2}{2m}\mathit{\Delta} + \sum_{\alpha}U_{C\alpha}^{(1)}
\label{h1}
\end{equation}
denote the \mbox{1-particle}~Hamiltonian. Suppose~$U_C$ is a sufficiently regular function 
of~$|\vec{R}_{\alpha}-\vec{r}|$, so that the integral kernel of~$\rho^{(1)}=\exp\left[ -\beta 
h^{(1)}\right]$ admits the~\mbox{Feynman}-\mbox{Kac} representation~\cite{simon,glimm}, viz., 
\begin{equation}
	\rho(\vec{r},\vec{r'})=\int_{\Omega_0}\mathrm{d}\mu_{\vec{r},\vec{r'}}(\mbox{\boldmath
	$\omega$})\exp\left[- \sum_{\alpha} \int_0^{\beta}U_{C\alpha}(\mbox{\boldmath $\omega$}(s))\mathrm{d} s 
	\right]
\label{rho}
\end{equation}
where~$U_{C\alpha}(\vec{r})=U_C(\vec{R}_{\alpha}-\vec{r})$. Due to randomness of ion 
positions~$\vec{R}_{\alpha}$, $\rho(\vec{r},\vec{r'})$ is equal to its space 
average~$\left\langle\rho(\vec{r},\vec{r'}) \right\rangle_{\Lambda}$ over these 
positions~\cite{edwards,ambegaokar}. In order to evaluate~$\left\langle \rho(\vec{r},\vec{r'}) 
\right\rangle_{\Lambda}$, let us note that~$U_{C\alpha}$ is negative-valued and has a finite 
minimum~\cite{ziman}. Suppose it admits a \mbox{Taylor}~expansion in the neighbourhood of this minimum 
at~$\vec{r}_{0\alpha}$ and that~$\beta$ is sufficiently large. Then
\begin{equation}
	\left\langle\rho(\vec{r},\vec{r'})\right\rangle_{\Lambda}=|\Lambda|^{-M} \prod_{\alpha=1}^M 
	\int_{\Lambda} \mathrm{d}^3 R_{\alpha} \rho(\vec{r},\vec{r'})
\label{rho_av}
\end{equation}
can be evaluated by expanding~$U_C (\vec{R}_{\alpha}-\vec{r})$ up to second order and applying
the method of steepest descent. Evaluation of the average~\eqref{rho_av} thus amounts to
calculating the integral
\begin{equation}
	\prod_{\alpha}\int_{\Lambda}\mathrm{d}^3 R_{\alpha}\exp\left[ -\beta u_0-\frac{1}{2} u_2
	\int_0^{\beta}\sum_{q=1}^3\left(R_{\alpha q}-r_{0\alpha q}-\omega_q(s)\right)^2\mathrm{d} s +\ldots\right]
	\label{expan}
\end{equation}
where~$u_0=U_C\left( \vec{R}_{\alpha}-\vec{r}_{0\alpha}\right)$, $u_2=U_C''\left( \vec{R}_{\alpha}-
\vec{r}_{0\alpha}\right)$. This is done in Appendix~\ref{a}. One obtains for sufficiently large~$\beta$
\begin{equation}
	\begin{split}
	&\left\langle\rho \left( \vec{r},\vec{r'}\right) \right\rangle_{\Lambda}=Z_1 \int_{\Omega_0}\mathrm{d}\mu_{\vec{r},
	\vec{r'}}\left( \mbox{\boldmath $\omega$} \right) \int_{\mathbb{R}^3} \exp \left[ -\beta M u_0 \right. \\
	&\left. -\frac{1}{2} M u_2 \sum_{q=1}^3 \int_0^{\beta}\left( \omega_q(s)-
	\lambda_q \right)^2 \mathrm{d} s \right] \mathrm{d} \lambda_1 \mathrm{d} \lambda_2 \mathrm{d} \lambda_3 \\
\end{split}
	\label{rho_av_2}
\end{equation}
where
\begin{equation*}
	Z_1=|\Lambda|^{-M}\left( \frac{2\pi}{\beta u_2}\right)^{\frac{3}{2}(M-1)} M^{3/2}
\end{equation*}
The integral kernel~\eqref{rho_av_2} thus represents, up to a constant factor, the kernel of the canonical
density operator of a particle oscillating around a point~$\vec{\lambda}$, integrated over all 
positions~$\vec{\lambda}\in \mathbb{R}^3$. In other words,
\begin{equation}
	|\Lambda|^{-M}\prod_{\alpha}\int_{\Lambda}\mathrm{d}^3 R_{\alpha} \rho^{(1)}=Z_1 \exp[-\beta M u_0] \int_{\mathbb{R}^3}\mathrm{d}^3 \lambda
	\exp \left[ -\beta h_{\mathrm{osc}}^{(1)}(\vec{\lambda})\right]
	\label{int}
\end{equation}
where
\begin{equation}
	h_{\mathrm{osc}}(\vec{\lambda})=-\frac{\hslash^2}{2m}\mathit{\Delta} +\frac{1}{2} Mu_2 \left( \vec{r}-\vec{\lambda}\right)^2
	\text{.}
	\label{h_osc}
\end{equation}
The integral kernel of~$\exp \left[ -\beta h_{\mathrm{osc}}^{(1)}(\vec{\lambda})\right]$ 
is~\cite{simon,glimm,pruski}
\begin{equation}
	\rho_{\mathrm{osc}}(\vec{r},\vec{r'})=\xi^3 \exp \left[ -\frac{1}{2} a \left[ \left(\vec{r}-
	\vec{\lambda}\right)^2+ 
	\left( \vec{r'}-\vec{\lambda}\right)^2\right]+b\left(\vec{r}-\vec{\lambda} 
	\right)\left(\vec{r'}-\vec{\lambda}\right)\right]
\label{rho_osc}
\end{equation}
with
\begin{equation*}
	\xi=\frac{\alpha\left( \tanh \frac{1}{2} \Omega \right)^{1/2}}{2\sqrt{\pi}\sinh
	\frac{1}{2}\Omega}\quad\text{,} \quad a=\alpha^2\coth \Omega \quad \text{,} \quad
	b=\frac{\alpha^2}{\sinh \Omega}\text{,}
\end{equation*}
\begin{equation*}
	\alpha^2=\hslash^{-1}\sqrt{Mmu_2}\quad \text{,} \quad
	\Omega=\frac{\hslash^2\alpha^2}{m}\beta\text{.}
\end{equation*}
The integral over $\mathbb{R}^3$ on the rhs of Eq.~\eqref{rho_av_2} is thus a product of three 
\mbox{1-dimensional} integrals
\begin{equation}
	\begin{split}
	&\int_{-\infty}^{\infty}\mathrm{d} \lambda \exp \left[ -\frac{1}{2} a \left[\left(x-\lambda \right)^2+
	\left(x'-\lambda \right)^2\right]+b\left(x-\lambda \right)\left(x'-\lambda \right)\right]\\
	&=\sqrt{\frac{\pi}{a-b}}\exp \left[ -\frac{1}{4} \left( a+b\right)\left( x-x'\right)^2\right] \\
\end{split}
\label{1dim_int}
\end{equation}
each of which is equal to the integral kernel of the operator~\cite{simon,glimm}
\begin{equation}
	\sqrt{\frac{\pi}{a-b}}\sqrt{2\pi t_0}\exp\left[ -t_0 T_0\right]=\frac{2\pi}{\sqrt{a^2-b^2}}\exp
	\left[ -t_0 T_0 \right]\text{,}
\label{1dim_int_2}
\end{equation}
where
\begin{equation*}
	t_0=\frac{2}{a+b}=2\alpha^{-2}\tanh \frac{1}{2}\Omega \quad\text{,}\quad T_0=- \frac{1}{2} 
	\frac{\mathrm{d}^2}{\mathrm{d} x^2}\text{.}
\end{equation*}
One finds
\begin{equation}
	t_0 T_0=t(u_2,\beta)\left( -\frac{\hslash^2}{2m}\frac{\mathrm{d}^2}{\mathrm{d}
	x^2}\right)=t(u_2,\beta)H_{0x}^{(1)}
	\label{t0}
\end{equation}
with 
\begin{equation*}
	t(u_2,\beta)=\delta^{-1}\tanh (\delta\beta)\quad\text{,}\quad \delta=\frac{\hslash^2}{2m}\alpha^2
	\text{.}
\end{equation*}
Combining Eqs.~\eqref{rho_av_2},~\eqref{1dim_int},~\eqref{1dim_int_2},~\eqref{t0}, one obtains for 
large~$\beta$ the asymptotic equality
\begin{equation}
	\left\langle\rho(\vec{r},\vec{r'})\right\rangle_{\Lambda}=Z_0(u_0,u_2)\left( 2\pi t_0\right)^{-
	3/2} \exp \left[-\frac{1}{2t_0}\left( \vec{r}-\vec{r'}\right)^2\right]\text{,}
\label{rho_av_3}
\end{equation}
where
\begin{equation*}
	Z_0(u_0,u_2)=|\Lambda|^{-M}\left( \frac{2\pi}{\beta u_2}\right)^{\frac{3}{2} (M-1)}M^{3/2} \xi^3 
	\left(\frac{2\pi}{\alpha^2}\right)^3 \exp \left[-\beta Mu_0\right]\text{.}
\end{equation*}
Thus
\begin{equation}
	\left\langle \rho^{(1)}\right\rangle_{\Lambda}=Z_0\exp\left[ -tH_0^{(1)}\right]
\label{rho_av_4}
\end{equation}
In the low-temperature r\'egime, $\left\langle \rho^{(1)}\right\rangle_{\Lambda}$~is thus equal, up to a 
normalization factor, to the canonical density operator of a free particle at an effective 
temperature~$\left( \mathrm{k}_{\mathrm{B}}t\right)^{-1}>\left( \mathrm{k}_{\mathrm{B}}\beta\right)^{-1}=T$.
\section{\mbox{Mean-field} impurity heat capacity}\label{3}
Formulae for \mbox{s-d} system's \mbox{mean-field} energy~$\Delta U_{\mathrm{s-d}}$ and heat capacity~$\Delta 
C$, relative to that of free electrons, will be now derived, using the Hamiltonian~$h^{(n,M)}$, and compared 
with experimental data on impurity heat capacity of~$\mathrm{CuCr}$~\cite{triplett}
and~$\mathrm{LaCe}\mathrm{Al}_2$~\cite{bader}. 
\subsection{$\mathrm{CuCr}$ alloys}
According to \mbox{Monod et al.}~\cite{monod} the spin of the~$\mathrm{Cr}^{3+}$~ions in~$\mathrm{CuCr}$
alloys equals~$3/2$. The impurity expectation energy in these alloys therefore equals 
\begin{equation}
	\begin{split}
		&U_{\mathrm{imp}}=\left\langle h_{\mathrm{imp}}^{(M)}\right\rangle_{h_{\mathrm{imp}}^{(M)}}\\
	&=-Mn\xi f_2(\xi)+\frac{1}{2} Mg^2+4Mg^2 \frac{\mathrm{e}^{-4\beta g^2}\cosh \left( 3\beta g \xi\sqrt{n} 
	\right)}{\mathrm{e}^{-4\beta g^2}\cosh\left( 3\beta g\xi\sqrt{n}\right)+\cosh\left( \beta g\xi\sqrt{n} 
	\right)}\text{.} \\
\end{split}
\label{U_imp}
\end{equation}
$f_1$, defined by Eq.~\eqref{f1} is a linear function in the simplest approximation~\eqref{f1_approx}:
$f_1(\xi)=b_0+b_1\xi$ (Appendix~\ref{b}). Then according to Eqs.~\eqref{b6},~\eqref{b8}, the interaction 
energy of electrons with the Hamiltonian~$\widetilde{h}_{\mathrm{e}}^{(n,M)}(\xi,\eta)$ equals
\begin{equation}
	\begin{split}
	\Delta U_{\mathrm{e}} &=\left\langle \widetilde{h}_{\mathrm{e}}^{(n,M)}(\xi,\eta)\right 
	\rangle_{\widetilde{h}_{\mathrm{e}}^{(n,M)}(\xi,\eta)}-\left\langle A^{(n)} H_0^{(n)} 
	\right\rangle_{A^{(n)} H_0^{(n)}}\\
	&=-Mnf_2(\xi)\left(b_0+b_1 f_2(\xi)+\ldots\right)\text{.}\\
	\end{split}
\label{deltaU}
\end{equation}

>From Eqs.~\eqref{h_nm},~\eqref{U_imp},~\eqref{deltaU} one obtains the following expression 
for~$\Delta U_{\mathrm{s-d}}$ of a system of $n$~electrons and $M$~impurities with spin~$3/2$:
\begin{equation}
	\Delta U_{\mathrm{s-d}}=U_{\mathrm{imp}}-Mnf_2(\xi)\left( b_0+b_1 f_2(\xi)+\ldots\right)+Mn\xi f_2(\xi)-\frac{1}{2}
	Mnf_2^2(\xi)
\label{deltaU_sd}
\end{equation}
where~$\xi$ is the minimizing solution of Eq.~\eqref{xi} with~$f_1$ given by Eq.~\eqref{f1_approx}. 

The number of conduction electrons per host atom in~$\mathrm{CuCr}$ equals~one, and hence the number of these 
electrons per impurity~$n_1=c^{-1}$.

The $n$-electron, $M$-impurity \mbox{s-d} system will be now treated as a subsystem of a sample $S$ 
containing one mole of impurities. Then~$n=Mn_1$. In terms of~$n_1$, $\gamma=\sqrt{M}g$, the energy~$\Delta
U_m=6.022\cdot 10^{23}\,M^{-1}\Delta U_{\mathrm{s-d}}$ of such sample with spin~$3/2$ impurities, divided 
by~$c$ and expressed in joules, equals
\begin{equation}
\begin{split}
	& c^{-1}\Delta U_{\mathrm{m}}=\left(\frac{1}{2}\gamma^2 + 4\gamma^2\frac{\mathrm{e}^{-4\beta\gamma^2 M^{-1}} 
	\cosh\left(3\beta\gamma\xi\sqrt{n_1}\right)}{\mathrm{e}^{-4\beta\gamma^2 M^{-1}}\cosh
	\left(3\beta\gamma\xi\sqrt{n_1}\right)+\cosh \left(\beta\gamma\xi\sqrt{n_1}\right)}\right.\\
	&\left. -\frac{1}{2} M^2 n_1 f_2^2(\xi)-M^2 n_1 f_2(\xi)\left( b_0+b_1f_2(\xi)+\ldots\right)\right) 
	602.2\cdot 160.2\, M^{-1}n_1 \\
\end{split}
\label{deltaUm}
\end{equation}
with
\begin{equation*}
	f_2(\xi)=\frac{\gamma}{M\sqrt{n_1}} \frac{3\mathrm{e}^{-4\beta\gamma^2 M^{-1}} 
	\sinh\left(3\beta\gamma\xi\sqrt{n_1}\right)+\sinh\left(\beta\gamma\xi\sqrt{n_1}\right)}{\mathrm{e}^{-
	4\beta\gamma^2 M^{-1}}\cosh\left(3\beta\gamma\xi\sqrt{n_1}\right)+
	\cosh\left(\beta\gamma\xi\sqrt{n_1}\right)}
\end{equation*}
if~$\gamma$,~$b_0$ are given in~$\sqrt{\mathrm{eV}}$.

Eq.~\eqref{xi} takes the form
\begin{equation}
	f_3(\xi)=\xi
	\label{f3}
\end{equation}
with~$f_3=b_0+\left( b_1+1\right)f_2$.

The excess heat capacity of~$S$, relative to that of pure $\mathrm{Cu}$, equals
\begin{equation}
	\Delta C(T)=\frac{\partial \Delta U_{\mathrm{m}}}{\partial T}+\frac{\partial \Delta
	U_{\mathrm{m}}}{\partial \xi} \frac{\partial \xi}{\partial T}\text{,}
\label{delatCT}
\end{equation}
where
\begin{equation}
	\frac{\partial \xi}{\partial T}=-\frac{\partial f_3}{\partial T}\left( \frac{\partial f_3}{\partial
	\xi}-1\right)^{-1}\text{.}
\label{partial}
\end{equation}
$\Delta C(T)$ depends on the solution~$\xi$ of Eq.~\eqref{f3} and on the parameters $b_0$, $b_1$,
$\gamma$,$M$. An additional parameter~$\Delta T>0$, equal to the shift of the temperature variable 
of~$\Delta C$, proves necessary. The origins of this shift are explained in Section~\ref{2}.

\begin{figure}
	\resizebox{0.85\columnwidth}{!}{\includegraphics{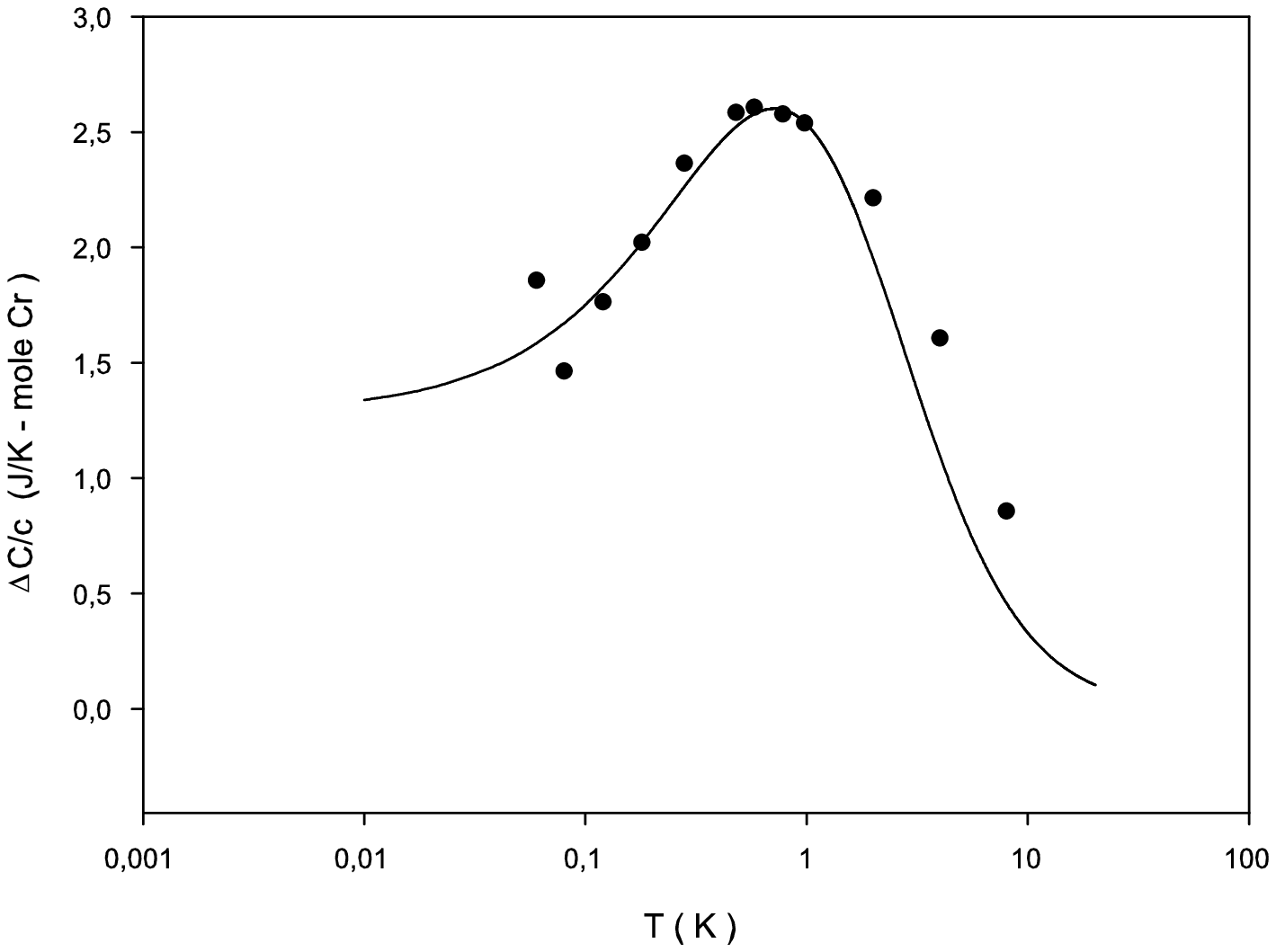}}	  
	\includegraphics{fig1.eps}
\caption{The \mbox{mean-field} impurity heat capacity of~$\mathrm{CuCr}$ with~$c=51\,\text{ppm}$
	and values of $b_0$, $b_1$, $\gamma$, $M$, $\Delta T$ given in Table~\ref{tab1}. The points are
	experimental results from Ref.~\cite{triplett}.}
\label{fig1}
\end{figure}

\begin{figure}
	\resizebox{0.85\columnwidth}{!}{\includegraphics{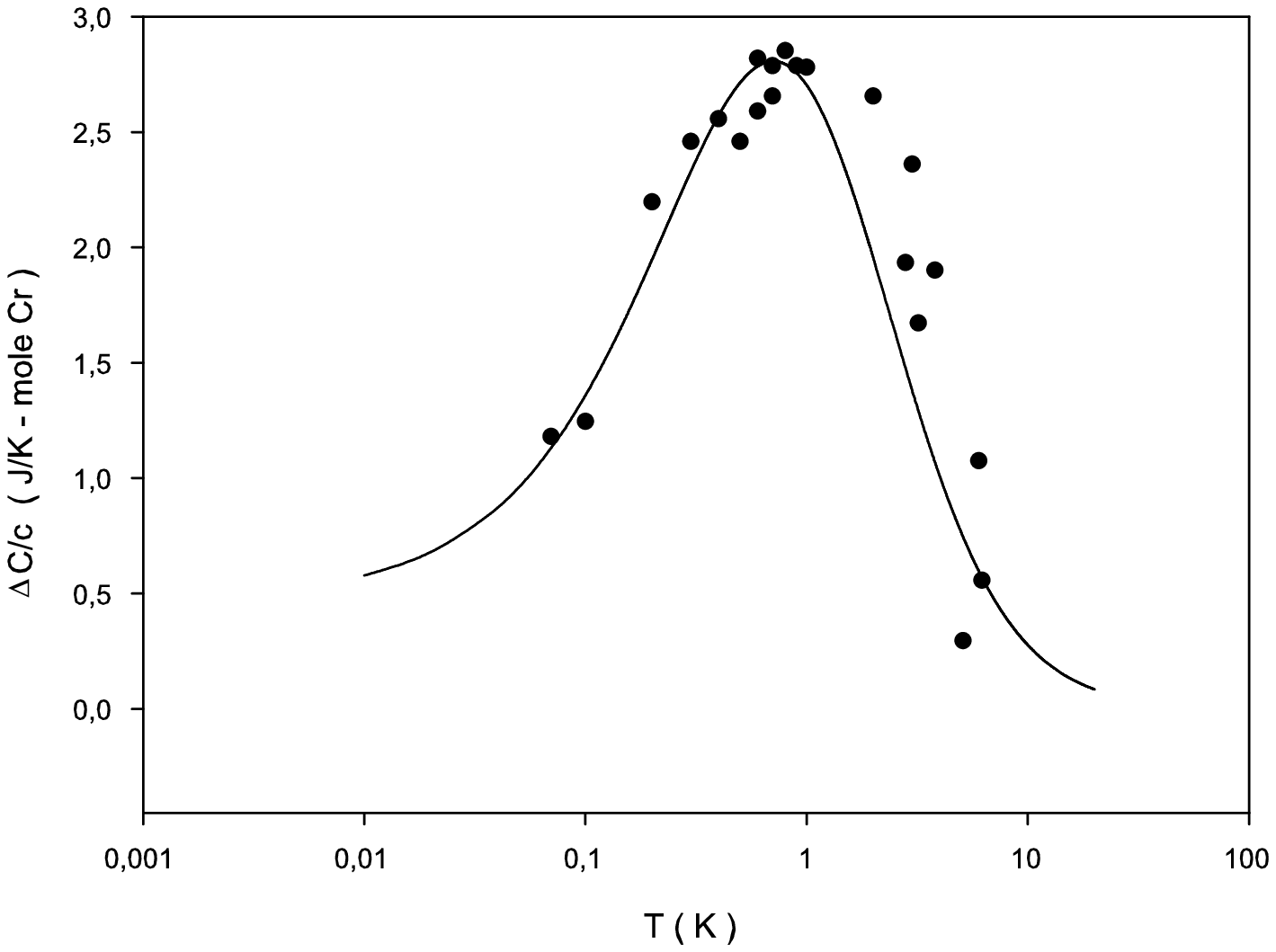}}
	\includegraphics{fig2.eps}
	\caption{The \mbox{mean-field} impurity heat capacity of~$\mathrm{CuCr}$ with~$c=21.2\,\text{ppm}$
        and values of $b_0$, $b_1$, $\gamma$, $M$, $\Delta T$ given in Table~\ref{tab1}. The points
	are experimental results from Ref.~\cite{triplett}.}
\label{fig2}
\end{figure}

The best fitting graphs of~$c^{-1}\Delta C(T+\Delta T)$ vary with~$M$. Moreover, the graphs obtained 
for~$M\gg 1$ provide much better agreement with experiment than those for~$M=1, 2$. The graphs 
of~$c^{-1}\Delta C(T+\Delta T)$ for~$\mathrm{CuCr}$ with~$c=51\,\text{ppm}$ ($M=24850$) and~$c=21.2\,
\text{ppm}$ ($M=360000$), plotted in the same units as in Ref.~\cite{triplett}, are depicted in 
Figs.~\ref{fig1},~\ref{fig2}. The corresponding values of the remaining parameters $b_0$, $b_1$, $\gamma$,
$\Delta T$ are given in Table~\ref{tab1}. 

\begin{table}
\begin{tabular}{ccccccccc}
	\hline
	Alloy & $x$ & $c$ & $n_1$ & $b_1$ & $b_0\, [\sqrt{\mathrm{eV}}]$ & $\gamma\, [\sqrt{\mathrm{eV}}]$ &
	$M$ & $\Delta T\, [\mathrm{K}]$\\ \hline
	$\mathrm{CuCr}$ & - & $\frac{51}{10^6}$ & $\frac{10^6}{51}$ & $-461$ & $1.09\cdot\frac{10^{-3}}{n_1}$ 
	& $0.091$ & $248500$ & $1.05$ \\\hline
	$\mathrm{CuCr}$ & - & $\frac{212}{10^7}$ & $\frac{10^7}{212}$ & $-631$ &
	$1.01\cdot\frac{10^{-3}}{n_1}$ & $0.086$ & $36\cdot 10^4$ & $0.78$ \\\hline
	$\left( \mathrm{LaCe}\right)\mathrm{Al}_2$ & $\frac{64}{10^4}$ & $\frac{4}{1871}$ & $\frac{3742}{3}$ 
	& $-13101$ & $\frac{19}{10^{5}}$ & $\frac{31}{10^5}$ & $58$ & $0.39$ \\\hline
\end{tabular}
\caption{}
\label{tab1}
\end{table}

\begin{figure}
	\resizebox{0.85\columnwidth}{!}{\includegraphics{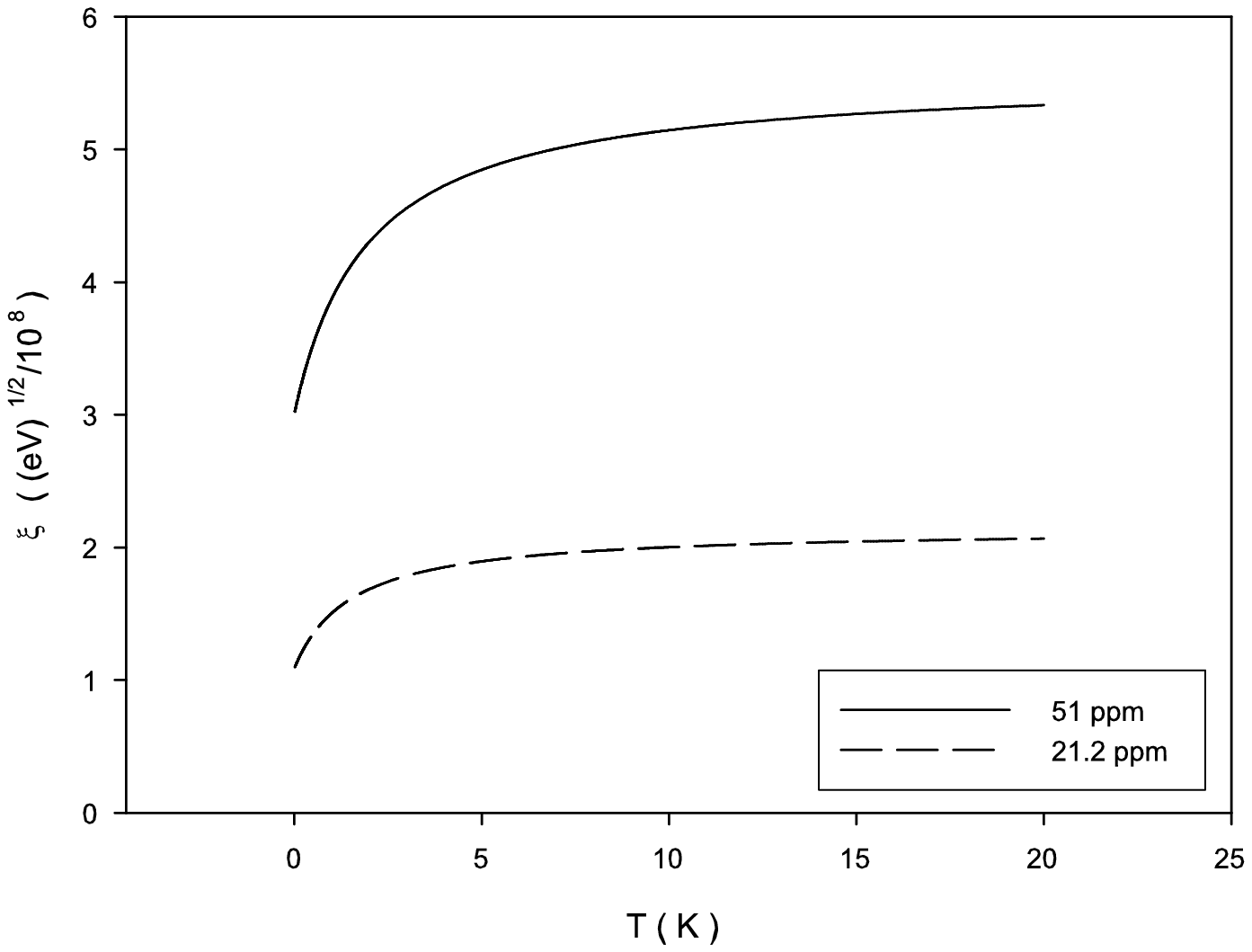}}	  
	\includegraphics{fig3.eps}
\caption{The solution~$\xi(T)$ of Eq.~\eqref{f3} for~$\mathrm{CuCr}$ with values of $b_0$, $b_1$, $\gamma$, 
$M$, $\Delta T$ given in Table~\ref{tab1}.}
\label{fig3}
\end{figure}

The solution~$\xi(T)$ of Eq.~\eqref{f3} is unique in both cases and minimizes $f(h^{(n,M)},\beta)$. The graphs of~$\xi(T)$ are plotted in Fig.~\ref{fig3}. 

It appears that higher order terms of the expansions~\eqref{b6},~\eqref{f1_approx} should be included in 
order to improve the quality of \mbox{mean-field}~$\Delta C/c$.

It is worth noting that since the best fitting values $M_f$ of $M$ are much smaller than $A=6.022\cdot
10^{23}$, the sample $S$ can be viewed as consisting of magnetic domains, each containing $M_f$ impurities
with a definite favoured impurity-spin orientation, which differs, in general, from one domain to another.
Existence of such domains in some magnetic materials has been established experimentally
(e.g.~Ref.~\cite{aharoni}). 
\subsection{$\left( \mathrm{La}_{1-x}\mathrm{Ce}_x\right)\mathrm{Al}_2$}
Experimental data on~$\Delta C/c$ of~$\left( \mathrm{La}_{1-x}\mathrm{Ce}_x\right)\mathrm{Al}_2$ alloys are 
presented in Ref.~\cite{bader}. According to Refs.~\cite{bader,felsch}, a typical \mbox{Kondo}~effect, 
without any superconducting side-effects, is observed in~$\left(
\mathrm{La}_{1-x}\mathrm{Ce}_x\right)\mathrm{Al}_2$ samples with~$\mathrm{Ce}$ content above~$x=0.0067$. 
However, according to \mbox{Bader et al.}~\cite{bader}, for~$x=0.0064$ the expected normal-state
and measured superconducting-state heat capacities do not differ significantly. Thus a \mbox{mean-field}
normal-state theory of~$\Delta C/c$ for~$\left( \mathrm{La}_{1-x}\mathrm{Ce}_x\right)\mathrm{Al}_2$ 
with~$x=0.0064$ can be reliable.

The number of valence electrons per host atom in~$\mathrm{LaAl}_2$ equals~$8/3$. For a~$x=0.0064$, 
\begin{equation*}
	c=\frac{0.0064}{2.9936}=\frac{4}{1871}\quad \text{,} \quad
	n_1=\frac{8}{3}c^{-1}=\frac{3742}{3}\text{.}
\end{equation*}
Since the spin of~$\mathrm{Ce}$ ions equals~$1/2$~\cite{bader}, therefore
\begin{equation}
	f_2(\xi)=\frac{\gamma}{M\sqrt{n_1}}\tanh\left( \beta \gamma\xi\sqrt{n_1}\right)
	\label{f2_La}
\end{equation}
and~$\Delta U_{\mathrm{m}}/c$ for a sample~$S$ of~$\left( \mathrm{La}_{1-x}\mathrm{Ce}_x\right) 
\mathrm{Al}_2$, expressed in joules, equals
\begin{equation}
	\begin{split}
		c^{-1}\Delta U_{\mathrm{m}} &= \left( \frac{1}{2} \gamma^2 -\frac{1}{2} M^2 n_1 f_2^2(\xi)-M^2 n_1 f_2(\xi) \left[ b_0
	\right. \right. \\
	& \left. \left. + b_1 f_2(\xi)+\ldots\right] \right) \frac{1}{4}\cdot 602.2 \cdot 160.2\cdot 1871 M^{-1} \\
\end{split}
\label{DeltaUm}
\end{equation}
if Eq.~\eqref{b8} is used and $b_0$, $\gamma$, $\beta$, $\xi$ are given in powers of~$\mathrm{eV}$.

\begin{figure}
	\includegraphics{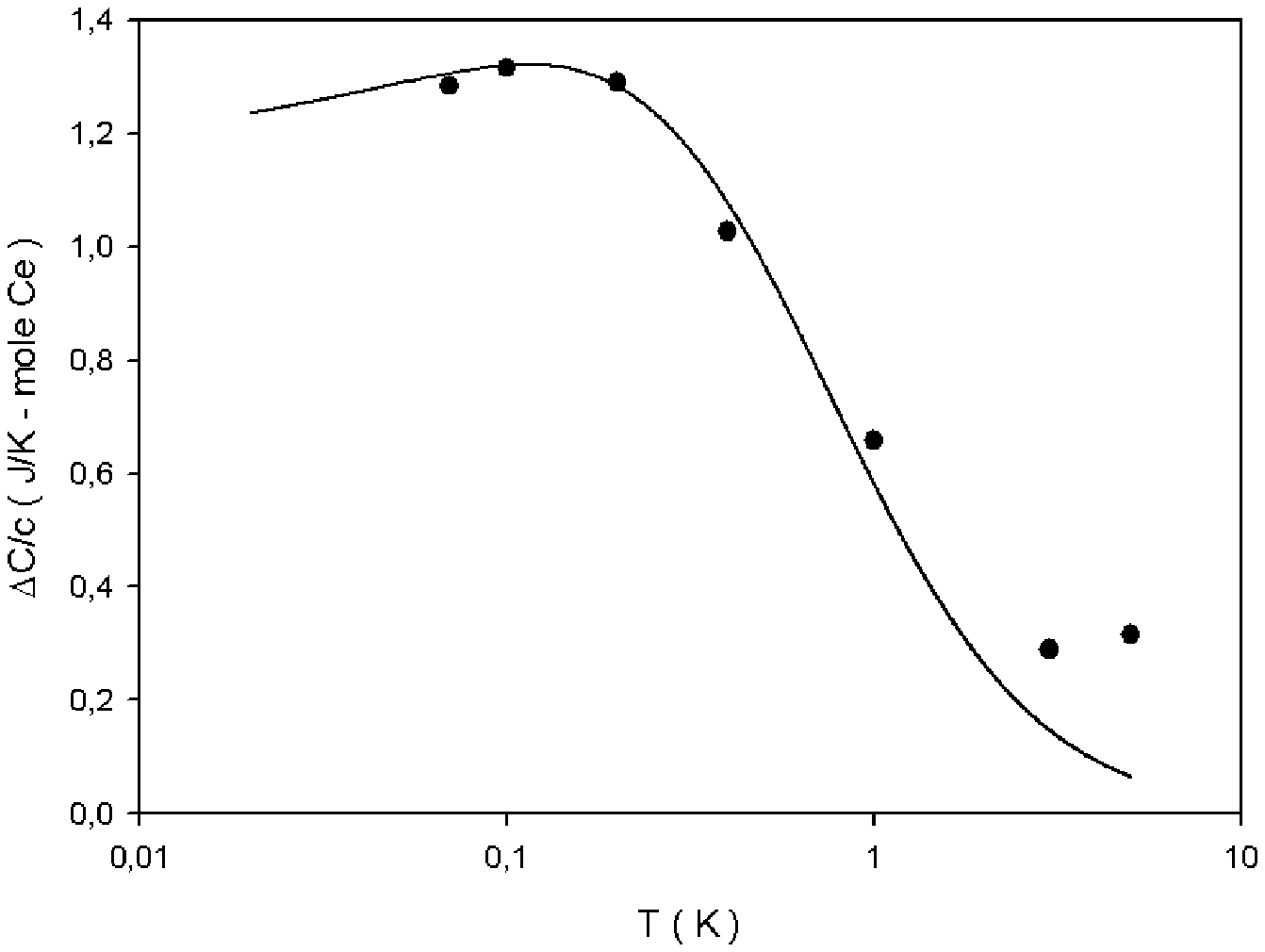}
\caption{The \mbox{mean-field} impurity heat capacity of~$\left(
\mathrm{La}_{1-x}\mathrm{Ce}_x\right)\mathrm{Al}_2$ with~$x=0.0064$ and values of $b_0$, $b_1$,
$\gamma$, $M$, $\Delta T$ given in Table~\ref{tab1}. The points are experimental results from
Ref.~\cite{bader}.}
\label{fig4}
\end{figure}

The \mbox{mean-field}~$\Delta C/c$ curve best fitting to experimental~$\Delta C/c$ of Ref.~\cite{bader} was 
obtained for~$M=40$ and is depicted in Fig.~\ref{fig4}. The corresponding values of other parameters are 
given in Table~\ref{tab1}. Since the error of experimental~$\Delta C/c$ values is relatively high 
above~$5\,\mathrm{K}$~\cite{bader}, the \mbox{mean-field}~$\Delta C/c$ curve in Fig.~\ref{fig4} provides a 
good fit to experiment. 

The equation~$f_3(\xi)=\xi$ for~$\xi(T)$ has a unique minimising solution with~$\xi(T)\in \left( 11\cdot
10^{-5}\,\sqrt{\mathrm{eV}}, 19\cdot 10^{-5}\,\sqrt{\mathrm{eV}}\right)$ for~$T\in\left(
0.015\,\mathrm{K}, 10\,\mathrm{K}\right)$. The graph of~$\xi(T)$ is, similar as those for~$\mathrm{CuCr}$ in 
Fig.~\ref{fig3}, increasing in~$T$.
\section{Concluding remarks}\label{4}
Progress has been made in improving the quality of theoretical impurity heat capacity graphs~$\Delta
C(T)$ of~DMA. Scaling procedures, used in previous \mbox{s-d}~theories to adjust thermodynamic functions to 
experimental data, were unnecessary and nonlinear dependence of~$\Delta C(T)$ on impurity concentration
$c$ in $\mathrm{CuCr}$ has been accounted for the first time.

The \mbox{mean-field} theory of a dilute \mbox{s-d}~system established in Ref.~\cite{m2} has also proved 
capable of explaining the temperature dependence of impurity heat capacity~$\Delta C$ of~$\left(
\mathrm{La}_{1-x}\mathrm{Ce}_x\right)\mathrm{Al}_2$ with~$x=0.0064$. Partial quantitive agreement between 
theory and experiment has been achieved for~$\mathrm{CuCr}$ alloys, especially for the smaller value of impurity concentration considered.

Higher order terms of the expansion of~$f_1$ should be included in order to improve these 
results. 
\appendix
\section{}\label{a}  
The integral~\eqref{expan} is an~$M$-fold product of~3-dimensonal integrals
\begin{equation}
	I=\int_{\Lambda}\mathrm{d}^3 R \exp \left[ -\frac{1}{2} u_2 \int_0^{\beta}\sum_{q=1}^3 \left(
	R_q-r_{0q}-\omega_q(s)\right)^2 \mathrm{d} s \right]
	\label{I}
\end{equation}
and~$\exp \left[ -\beta Mu_0\right]$.  The representation~\eqref{rho_av_2} of~$\left\langle \rho(\vec{r},
\vec{r'})\right\rangle_{\Lambda}$, for large~$\beta$, obtains by apllying to the~$I$ the dominated 
convergence theorem (in order to express the integral in the exponent as a limit of partial sums) and the 
method of steepest descent:
\begin{equation}
	\begin{split}
	&I=\lim_{m\to\infty} \int_{\Lambda} \mathrm{d}^3 R \exp \left[ -\frac{\beta}{2m}u_2 \sum_{k=1}^{m}\sum_q
	\left( R_q-r_{0q}-x_{kq} \right)^2 \right] \\
	& \approx \lim_{m\to\infty} \int_{\mathbb{R}^3} \mathrm{d}^3 R \exp
	\left[ -\frac{\beta}{2m}u_2 \sum_{kq} x_{kq}^2 \right. \\
	&\left. - \frac{\beta}{2m}u_2 \sum_{q} \left( \frac{1}{\sqrt{m}}\sum_k x_{kq}-\sqrt{m} \left( R_q -
	r_{0q}\right)\right)^2 + \frac{\beta}{2m^2}u_2 \sum_{q}\left( \sum_k x_{kq} \right)^2 \right] \\
	& = \lim_{m\to\infty} \left( \frac{2\pi}{\beta u_2}\right)^{3/2} \exp \left[ -\frac{\beta}{2m} u_2 
	\sum_{kq} x_{kq}^2 +\frac{\beta}{2m^2} u_2 \sum_q \left( \sum_k x_{kq} \right)^2 \right] \\
	& = \left( \frac{2\pi}{\beta u_2}\right)^{3/2} \exp \left[ -\frac{1}{2} u_2
	\int_0^{\beta} \mbox{\boldmath{$\omega$}}^2(s) \mathrm{d} s + \frac{1}{2} \beta^{-1} u_2 \sum_q 
	\left( \int_0^{\beta} w_q(s) \mathrm{d} s 
	\right)^2 \right]\text{,} \\
\end{split}
\label{integral}
\end{equation}
where~$x_{kq}=\omega_q\left(\frac{k\beta}{m}\right)$. The squared integral in the exponent can be linearized,
using the identity
\begin{equation*}
	\exp(a^2)=(2\pi)^{-1/2} \int_{-\infty}^{\infty} \exp \left[ -\frac{1}{2} \xi^2 +\sqrt{2} a \xi\right] \mathrm{d} \xi
\end{equation*}
which yields
\begin{equation}
	\begin{split}
	& I^M=\left( \frac{2\pi}{\beta u_2}\right)^{3M/2} \exp \left[ -\frac{1}{2} M u_2 \int_0^{\beta}
	\mbox{\boldmath{$\omega$}}^2(s) \mathrm{d} s + \frac{1}{2} \beta^{-1} M u_2 \sum_q \left( \int_0^{\beta} w_q(s) \mathrm{d} 
	s \right)^2 \right] \\
	&= (2\pi)^{-3/2} \left( \frac{2\pi}{\beta u_2}\right)^{3M/2} \int_{\mathbb{R}^3} \mathrm{d} \xi_1 \mathrm{d} \xi_2 
	\mathrm{d} \xi_3 \exp \left[ -\frac{1}{2} \sum_s \xi_s^2 \right. \\
	&\left. - \frac{1}{2} M u_2 \int_0^{\beta} \mbox{\boldmath{$\omega$}}^2(s) \mathrm{d} s + \sqrt{\beta^{-1} M u_2} 
	\sum_q \xi_q \int_0^{\beta} \omega_q (s) \mathrm{d} s \right] \\
	& = \left( \frac{2\pi}{\beta u_2}\right)^{\frac{3}{2} (M-1)} M^{3/2} \int_{\mathbb{R}^3} \mathrm{d} \lambda_1 \mathrm{d} 
	\lambda_2 \mathrm{d} \lambda_3 \exp \left[ -\frac{1}{2} M u_2 \sum_q \int_0^{\beta} \left( \omega_q(s)-\lambda_q 
	\right)^2 \mathrm{d} s \right] \\
\end{split}
\label{IM}
\end{equation}
in accord with Eq.~\eqref{rho_av_2}.
\section{}\label{b}
The form of~$f_1$ defined by Eq.~\eqref{f1} is crucial in the \mbox{mean-field} theory of the 
\mbox{s-d}~system. Here~$f_1$ is derived under the assumption that~$\widetilde{h}_{\mathrm{e}}^{(1,M)}(\xi,
\eta)$ is defined in terms of a function~$-g\sqrt{n} U\left( |\vec{R}_{\alpha}-\vec{r}|\right)$, which 
has absolute maximum equal~$v_0$ and absolute minimum equal~$u_0$, and that the expansion in~\eqref{expan} is 
terminated on the first term. The space average of~$\exp \left[ -\beta \widetilde{h}_{\mathrm{e}}^{(1,M)}(\xi,
0)\right]$ over impurity positions then equals
\begin{equation*}
	\left\langle \exp \left[ -\beta \widetilde{h}_{\mathrm{e}}^{(1,M)}(\xi,0)\right] \right 
	\rangle_{\Lambda} = 
	\begin{cases}
		\exp \left[ -\beta H_0^{(1)} -\beta M u_0 \xi \right] & \text{for} \sigma_z=1 \\
		\exp \left[ -\beta H_0^{(1)} +\beta M v_0 \xi \right] & \text{for} \sigma_z=-1 \\
	\end{cases}
\end{equation*}
and Eq.~\eqref{z} for~$z$ takes the form
\begin{equation}
	\begin{split}
	\frac{4\pi v}{h^3} \int_0^{\infty} \mathrm{d} p\, p^2 & \left\{ \left( z^{-1} \exp \left[ \beta \left( M
	u_0 \xi+ \frac{p^2}{2m}\right)\right] + 1 \right)^{-1} \right. \\
	& \left. + \left( z^{-1} \exp \left[ \beta \left( -M v_0 \xi+ \frac{p^2}{2m}\right)\right] + 1 
	\right)^{-1} \right\} = 1\text{,} \\
\end{split}
\label{z_int}
\end{equation}
where~$v = |\Lambda| n^{-1}$. Following \mbox{Huang}~\cite{huang}, one transforms Eq.~\eqref{z_int} to 
\begin{equation}
	v \lambda_0^{-3} \left( f_{3/2}\left(z\mathrm{e}^{-\beta M u_0 \xi}\right) + f_{3/2}\left(z\mathrm{e}^{\beta M v_0
	\xi}\right) \right) = 1 \text{,}
	\label{f32}
\end{equation}
where 
\begin{equation*}
	v \lambda_0^{-3} = \frac{1}{8} 3 \sqrt{\pi} \left( \beta \varepsilon_{\mathrm{F}} \right)^{-3/2} 
	\text{.}
\end{equation*}
For large~$z$, $f_{3/2}(z)$~is given by \mbox{Sommerfeld's}~expansion:
\begin{equation}
	f_{3/2}(z) = \frac{4}{3\sqrt{\pi}} \left( \left( \ln z \right)^{3/2} + \frac{\pi^2}{8} \left( \ln z
	\right)^{-1/2} + \ldots \right)\text{.}
	\label{f32_exp}
\end{equation}
Let~$M u_0 \xi \varepsilon_{\mathrm{F}}^{-1} = \mu$, $M v_0 \xi \varepsilon_{\mathrm{F}}^{-1} = \nu$. 
Then~$z$ satisfying Eq.~\eqref{f32} is given by the expansion
\begin{equation}
	\ln z = a_1 \beta \varepsilon_{\mathrm{F}} + a_0 + a_{-1} \left( \beta \varepsilon_{\mathrm{F}} 
	\right)^{-1} + \ldots 
\label{lnz}
\end{equation}
with
\begin{equation*}
	a_1(\xi) = 1 + \frac{1}{2} \left( \mu - \nu \right) - \frac{1}{16} \left( \mu + \nu \right)^2 + \ldots 
\end{equation*}
\begin{equation*}
	a_0 = 0 
\end{equation*}
\begin{equation*}
	a_{-1}(\xi) = -\frac{\pi^2}{12} \left( 1 - \frac{1}{4} \left( \mu + \nu \right)^2 \right)^{-1/2} =
	-\frac{\pi^2}{12} \left( 1+ \frac{1}{8} \left( \mu + \nu \right)^2 + \ldots \right) \text{.}
\end{equation*}
>From Eq.~\eqref{f1} one finds (cf.\ Ref.~\cite{m2})
\begin{equation*}
	f_1(\xi) = \left( M n \beta \right)^{-1} \frac{\partial}{\partial \xi} \tr\ln \left( 1 + z
	\rho^{(1)} \right) \text{.}
\end{equation*}
Hence 
\begin{equation}
	\begin{split}
	f_1(\xi) &= v \lambda_0^{-3} \left( -u_0 f_{3/2}\left( z \mathrm{e}^{-\beta M u_o \xi}\right) + v_0
	f_{3/2}\left( z \mathrm{e}^{\beta M v_o \xi}\right) \right) \\
	& = b_0 + b_1 \xi -\frac{\pi^2}{8} b_0 \left( \beta \varepsilon_{\mathrm{F}} \right)^{-2} + \ldots \\
\end{split}
\label{f1_expand}
\end{equation}
with
\begin{equation*}
	b_0 = \frac{1}{2} \left( v_0 - u_0 \right) \quad \text{,} \quad b_1 = \frac{3M}{8\varepsilon_{\mathrm{F}}} 
	\left( u_0 + v_0 \right)^2 \text{.}
\end{equation*}
The energy of the electrons' subsystem also expresses in terms of~$f_1$. According to 
Eqs.~\eqref{h_e_tilde},~\eqref{f1}
\begin{equation}
	\begin{split}
		& \left\langle n \Gamma_1^n \widetilde{h}_{\mathrm{e}}^{(1,M)}(\xi, \eta) \right\rangle_{n 
		\Gamma_1^n \widetilde{h}_{\mathrm{e}}^{(1,M)}(\xi,\eta)}\\
		& =\left\langle n \Gamma_1^n H_0^{(1)} \right\rangle_{n \Gamma_1^n
		\widetilde{h}_{\mathrm{e}}^{(1,M)}(\xi,\eta)} - n M f_2(\xi) \left( b_0 + b_1 f_2(\xi) + 
		\ldots \right) \\
\end{split}
\label{b6}
\end{equation}
The first term on the rhs of Eq.~\eqref{b6} can be evaluated in a similar manner as the energy of a free
\mbox{Fermi}~gas, viz., 
\begin{equation}
	\begin{split}
		& \left\langle n \Gamma_1^n H_0^{(1)} \right\rangle_{n \Gamma_1^n
		\widetilde{h}_{\mathrm{e}}^{(1,M)}(\xi,\eta)}\\
		&= \frac{3n}{10} \varepsilon_{\mathrm{F}}^{-3/2} \beta^{-5/2} \left\{ \left( \ln z - \beta M 
		u_0 \left( \xi - \eta \right) \right)^{5/2} \right.\\
	&+ \frac{5}{8} \pi^2 \left( \ln z - \beta M u_0 \left( \xi - \eta \right)
	\right)^{1/2} + \ldots \\
	& \left. + \left( \ln z + \beta M v_0 \left( \xi - \eta \right) \right)^{5/2} + 
	\frac{5}{8} \pi^2 \left( \ln z + \beta M v_0 \left( \xi - \eta \right) \right)^{1/2} + \ldots
	\right\} \\
	& = \frac{3n}{10} \varepsilon_{\mathrm{F}} \left\{ 2 + 5 a_{-1}\left( \xi - \eta \right) \left( \beta
	\varepsilon_{\mathrm{F}} \right)^{-2} + \frac{5}{4}\pi^2 \left( \beta \varepsilon_{\mathrm{F}} 
	\right)^{-2} \right.\\
	&\left. + \frac{5}{8} M^2 \left( u_0 + v_0 \right)^2 \varepsilon_{\mathrm{F}}^{-2} \left( \xi - \eta 
	\right)^2 + \ldots \right\} \\
	& =  \left\langle n \Gamma_1^n H_0^{(1)} \right\rangle_{n \Gamma_1^n H_0^{(1)}} - \frac{\pi^2}{24} n M
	b_1\left(\xi - \eta \right)^2 \left( \beta \varepsilon_{\mathrm{F}} \right)^{-2} + \frac{1}{2} n M 
	b_1 \left(\xi - \eta \right)^2 + \ldots \\
\end{split}
\label{b7}
\end{equation}

If the \mbox{Taylor}~expansions of~$U_{\alpha}(\vec{r})$ around the absolute maximum and absolute minimum 
are cut off on the second derivatives, the structure of~$f_1$, as well as that of the expectation 
values~\eqref{b6},~\eqref{b7}, changes, e.g.\ under the approximations of Section~\ref{2}, $f_1$~is no longer 
regular at~$\xi = 0$:
\begin{equation*}
	\begin{split}
	f_1(\xi) & = v \left\{ \lambda_0 \left( t \left( u_2, \beta \right)\right)^{-3} \frac{\partial Z_0
	\left( u_0 \xi, u_2 \xi \right)}{\partial \xi} f_{3/2}\left( z Z_0 \left( u_0 \xi, u_2 \xi \right)
	\right) \right. \\
	& \left. + \lambda_0 \left( t \left( -v_2, \beta \right)\right)^{-3} \frac{\partial Z_0
	\left( -v_0 \xi, -v_2 \xi \right)}{\partial \xi} f_{3/2}\left( z Z_0 \left( -v_0 \xi, -v_2 \xi \right)
	\right) \right\} \text{,} \\
\end{split}
\end{equation*}
with
\begin{equation*}
	v \lambda_0^{-3}(t) = \frac{1}{8} 3\sqrt{\pi} \left( t \varepsilon_{\mathrm{F}} \right)^{-3/2}
\end{equation*}
($u_2$ ($v_2$) denoting the second derivative of~$-g\sqrt{n} U\left( \vec{R}_{\alpha} -\vec{r}\right)$ 
at the absolute minimum (maximum)) and the corresponding coefficient~$b_1'$, in the new expansion of~$f_1$ 
around~$\xi_0 \ne 0$, will differ from~$b_1''$ resulting in the modified Eq.~\eqref{b7}. In view of the 
presumable smallness of 
\begin{equation*}
	\Delta U_0 = \left\langle n \Gamma_1^n H_0^{(1)} \right\rangle_{\widetilde{h}_{\mathrm{e}}^{(n,
	M)}(\xi,\eta)} - \left\langle n \Gamma_1^n H_0^{(1)} \right\rangle_{n \Gamma_1^n H_0^{(1)}}
\end{equation*}
where~$\widetilde{h}_{\mathrm{e}}^{(n,M)} = n \Gamma_1^n \widetilde{h}_{\mathrm{e}}^{(1,M)}$, these modifications can be exploited 
in the simplest manner by restricting $u_0$, $v_0$, $u_2$, $v_2$ to values for which~$b_1''$ vanishes. As a
consequence,
\begin{equation}
	\Delta U_0 \approx 0
	\label{b8}
\end{equation}
whereas the range of~$b_1'$ can be expected to include also negative values. Thus, in general, 
\begin{equation}
	f_1(\xi)=b_0' + b_1' \xi + \ldots \quad \text{,} \quad b_0', b_1' \in \mathbb{R}^1
	\label{f1_approx}
\end{equation}
\bibliography{Mackowiak}
\bibliographystyle{unsrt}
\end{document}